\documentclass{revtex4}
\usepackage{amsfonts}
\usepackage{amsmath}
\usepackage{graphicx}
\usepackage{subfigure}

\begin{document}

\title{Spatiotemporal solitons in dispersion-managed multimode fibers}
\author{Thawatchai Mayteevarunyoo$^1*$, Boris A. Malomed$^{2,3}$, and Dmitry
V. Skryabin$^4$}

\affiliation{$^{1}$Department of Electrical and Computer Engineering, Faculty of
Engineering Naresuan University, Phitsanulok 65000, Thailand}
\affiliation{$^2$School of Electrical Engineering, Faculty of Engineering, Tel Aviv University, Tel
Aviv 69978, Israel}
\affiliation{$^3$Instituto de Alta Investigaci\'{o}n,
Universidad de Tarapac\'{a}, Casilla 7D, Arica, Chile}
\affiliation{$^{4}$Department of Physics, University of Bath, Bath, BA2 7AY, UK}

\begin{abstract}
We develop the scheme of dispersion management (DM) for three-dimensional
(3D) solitons in a multimode optical fiber. It is modeled by the parabolic
confining potential acting in the transverse plane in combination with the
cubic self-focusing. The DM map is adopted in the form of alternating
segments with anomalous and normal group-velocity dispersion. Previously,
temporal DM solitons were studied in detail in single-mode fibers, and some
solutions for 2D spatiotemporal `` light bullets", stabilized by DM, were
found in the model of a planar waveguide. By means of numerical methods, we
demonstrate that stability of the 3D spatiotemporal solitons is determined
by the usual DM-strength parameter, $S$: they are quasi-stable at $%
S<S_{0}\approx 0.93$, and completely stable at $S>S_{0}$. Stable vortex
solitons are constructed too. We also consider collisions between the 3D
solitons, in both axial and transverse directions. The interactions are
quasi-elastic, including periodic collisions between solitons which perform
shuttle motion in the transverse plane.
\end{abstract}
\maketitle

\section{Introduction}

Multidimensional solitons represent a vast research area comprising optics,
Bose-Einstein condensates (BECs) in ultracold gases, plasmas, liquid
crystals, and other areas \cite%
{early-review,early-review2,Akhmed,vortons,Ackemann,spatial-solitons,special-topics,Dumitru,NNR,Nature-Phys-reviews,Lavrentovich}%
. A fundamental problem is that the ubiquitous self-focusing cubic
nonlinearity, represented by the Kerr term in optics \cite{KA} or the
collisional one in the Gross-Pitaevskii equation for self-attractive BEC
\cite{Pit}, creates two- and three-dimensional (2D and 3D) solitons which
are unstable because the same cubic terms give rise to the critical and
supercritical collapse in 2D and 3D, respectively \cite{Gadi}. One
possibility for the stabilization of 3D matter-wave solitons in BEC,
including ones with embedded vorticity \cite{PhysD-vortices}, is the use of
the trapping parabolic, alias harmonic-oscillator (HO), potential \cite%
{Boris}, which is, in any case, a necessary ingredient in the experimental
realization of BEC \cite{Pit}. A qualitatively similar mechanism helps to
stabilize 3D `` optical bullets" \cite{Silberberg} (spatiotemporal solitons)
in multimode optical fibers, i.e., ones defined by the radial pattern of the
graded (refractive) index (GRIN) in the transverse cross-section plane. Such
fibers have been a subject of fundamental and applied research since long
ago \cite{MM1,MM2,MM3,MM4}, due to their potential for the use in optical
sensors \cite{sensors1,sensors2}, high-speed interconnects \cite%
{interconnects1,interconnects2}, and space-division multiplexing \cite%
{recentMM1,recentMM2,recentMM3}. The GRIN structure, supporting many
transverse modes (see, e.g., Ref. \cite{Eggleton1}), makes it possible to
consider 3D solitons as nonlinear superpositions of such modes, self-trapped
in the temporal dimension, i.e., along the fiber's axis \cite%
{Wang,Taiwan,Agrawal,Wabnitz,Agrawal2}. Recently, this approach to the study
of spatiotemporal solitons has drawn much interest \cite%
{sol-MM1,sol-MM2,sol-MM7,sol-MM3,sol-MM5,sol-MM4,instability,sol-MM6,laser,we}%
.

Another method for stabilization of both one- and multidimensional solitons,
in the form of oscillating breathers, with an intrinsic \textit{phase chirp}
\cite{KA}, is provided by "management" techniques. They are represented by
periodic modulation of basic parameters of the medium between positive and
negative values, along the propagation distance, in terms of optics, or in
time, in terms of BEC \cite{book}. A well-known example is the dispersion
management (DM), i.e., transmission of temporal solitons through a composite
line built as a concatenation of single-mode fibers with anomalous and
normal group-velocity dispersions (GVD) \cite{Turitsyn}. DM provides the
remarkable stabilization of the solitons in communication lines against
various perturbations, such as the Gordon-Haus jitter (induced by the
interaction of solitons with random optical radiation) \cite{GH}. Although
the DM\ soliton runs through the chain of segments with opposite values of
the GVD coefficient, which drives strong intrinsic oscillations in it,
extremely accurate simulations of the respective nonlinear Schr\"{o}dinger
equation (NLSE) have demonstrated that the ensuing oscillations of the
soliton's shape do not destabilize it, even if the propagation extends over
thousands of DM periods \cite{Nakazawa,Bennion,Nijhof,Liang}. Furthermore,
it was predicted that 2D spatiotemporal `` bullets" in a planar waveguide,
composed of alternating segments with opposite signs of GVD,\ also propagate
in the form of robust breathers with strong intrinsic oscillations \cite%
{Fatkhulla,Michal}. In particular, stable 2D breathers may feature
periodically recurring fission in two fragments and recombination into a
single soliton \cite{Michal}. In the bulk waveguide, 3D dispersion-managed
`` bullets" are unstable, but they may be stabilized by inclusion of the
defocusing quintic nonlinearity, which accounts for saturation of the cubic
self-focusing nonlinearity \cite{Wagner}. In addition to the stabilization
of solitons, the DM technique finds other applications to nonlinear optical
media, such as enhancement of supercontinuum generation \cite{Eggleton}.

While DM applies to optical media, the technique of `` nonlinearity
management", i.e., periodic alternation of self-focusing and defocusing, is
chiefly relevant to BEC, where it may be implemented by periodically
switching the sign of the nonlinearity with the help of the Feshbach
resonance controlled by an external magnetic field \cite{Inguscio,Randy},
which is made periodically time-dependent, for that purpose. The analysis,
performed in various forms, has predicted very efficient stabilization of 2D
ground-state breathers, while states with embedded vorticity and all 3D
solitons remain unstable under the action of the nonlinearity management
\cite{Towers,Fatkh,Ueda,Victor,SKA,JOSAB,Itin}. Matter-wave solitons may be
made stable in 3D if the time-periodic nonlinearity management is combined
with a quasi-1D spatially periodic potential (optical lattice) \cite{Michal2}%
.

The fact that DM\ helps to create and strongly stabilize oscillating
solitons in composite single-mode fibers \cite{Nijhof,book,Turitsyn}, and 2D
oscillatory `` bullets" in composite planar waveguides \cite%
{Fatkhulla,Michal}, suggests to consider a possibility of the creation of
robust spatiotemporal solitons in dispersion-managed multimode waveguides,
composed of alternating pieces of GRIN fibers with opposite signs of GVD. In
particular, the trend of the management to suppress instability against the
collapse \cite{book} offers a possibility to create stable high-power
solitons, that may be useful for applications.

Although splicing of segments of multimode fibers in the
composite system is a technological challenge, it has been implemented in
various forms, \cite%
{splicing-early,splicing-early-2,Horowitz,splicing-China,splicing-recent},
including large-transverse-area fibers with the same parabolic profile of
the local refractive index as considered in the present work \cite{splicing}.%
Alternative options are to replace one fiber species by a dispersive
grating \cite{grating}, or control the effective GVD by means of off-axis
light propagation \cite{off-axis}.

The objective of the present work is to identify stable and quasi-stable
spatiotemporal solitons in the DM\ multimode system, with the GRIN structure
represented by the HO trapping potential. We also address solitons with
embedded vorticity, as well as collisions between solitons.

The evolution of the complex amplitude $A\left( X,Y,Z,T\right) $ of the
electromagnetic field in the multimode fiber is governed by NLSE with
propagation distance $Z$, transverse coordinates $\left( X,Y\right) $, and
reduced time $T$ in the coordinate system traveling at the group velocity of
the carrier wave \cite{Anderson,Wang,Agrawal,sol-MM1,Kudl}:%
\begin{gather}
i\frac{\partial A}{\partial Z}=-\frac{1}{2k_{0}}\left( \frac{\partial ^{2}A}{%
\partial X^{2}}+\frac{\partial ^{2}A}{\partial Y^{2}}\right) +\frac{1}{2}%
\beta ^{\prime \prime }\left( Z\right) \frac{\partial ^{2}A}{\partial T^{2}}
\notag \\
+\frac{k_{0}\Delta }{R^{2}}\left( X^{2}+Y^{2}\right) A-\gamma \left\vert
A\right\vert ^{2}A,  \label{3DGPE}
\end{gather}%
where $k_{0}$ is the propagation constant of the carrier wave, and $\beta
^{\prime \prime }(Z)$ is the GVD coefficient which takes opposite values in
alternating segments of the multimode fibers. Further, the coefficient in
front of the HO potential, $\Delta \equiv \left( n_{\mathrm{core}}^{2}-n_{%
\mathrm{cladd}}^{2}\right) /\left( 2n_{\mathrm{core}}^{2}\right) $, is the
relative difference of the refractive index, $n$, between the fibers' core
and cladding, $R$ is the core's radius, and the nonlinearity coefficient is $%
\gamma \equiv k_{0}n_{2}/\left( n_{\mathrm{core}}\mathrm{A}_{\mathrm{eff}%
}\right) $, where $n_{2}$ is the Kerr coefficient, and $\mathrm{A}_{\mathrm{%
eff}}$ the effective area of the fiber's cross section. The present model
disregards the difference in $\gamma $ between different fiber segments, as
it is known that the DM is a much stronger factor than the difference
between different values of the self-focusing coefficient \cite{book}.

Equation (\ref{3DGPE}) is cast in a dimensionless form by the substitution:

\begin{gather}
x=X/w_{0},y=Y/w_{0},z=Z/\left( k_{0}w_{0}^{2}\right) ,  \notag \\
\tau =T/T_{0},u\left( x,y,\tau ;z\right) =\sqrt{\gamma k_{0}}w_{0}A\left(
X,Y,Z,T\right) ,  \label{scaling}
\end{gather}%
where $T_{0}$ and $w_{0}$ are the temporal and transverse scales (the
longitudinal one being $Z_{0}=k_{0}w_{0}^{2}$). This leads to the normalized
NLSE,

\begin{gather}
i{\frac{\partial u}{\partial z}}=-{\frac{1}{2}}\left( \frac{\partial ^{2}}{%
\partial x^{2}}+\frac{\partial ^{2}}{\partial y^{2}}\right) u-\frac{1}{2}D(z)%
\frac{\partial ^{2}u}{\partial \tau ^{2}}  \notag \\
+(x^{2}+y^{2})u-|u|^{2}u,  \label{3DGPE_nor}
\end{gather}%
where the transverse scale is chosen as $w_{0}=\left( R^{2}/k_{0}^{2}\Delta
\right) ^{1/4}$, to make the coefficient in front of the HO potential equal
to $1$, and $D(z)\equiv -\left( R/T_{0}^{2}\sqrt{\Delta }\right) \beta
^{\prime \prime }(Z)$.

The \textit{DM map}, i.e., the scheme of the periodic alternation of the GVD
coefficient in Eq. (\ref{3DGPE_nor}), is defined as follows:
\begin{equation}
D(z)=\left\{
\begin{array}{c}
D_{\mathrm{anom}}, \\
D_{\mathrm{norm}}, \\
D_{\mathrm{anom}},%
\end{array}%
\right.
\begin{array}{c}
0<z<\frac{1}{2}z_{\mathrm{anom}}, \\
\frac{1}{2}z_{\mathrm{anom}}\text{ }<z<\frac{1}{2}z_{\mathrm{anom}}+z_{%
\mathrm{norm}}, \\
\frac{1}{2}z_{\mathrm{anom}}+z_{\mathrm{norm}}<z<z_{\mathrm{map}}=0.5.%
\end{array}
\label{Disp_map}
\end{equation}%
Here, $D_{\mathrm{anom}}=\Delta D+D_{\mathrm{av}}>0$ and $D_{\mathrm{norm}%
}=-\Delta D+D_{\mathrm{av}}<0$ refer to the segments with the anomalous and
normal GVD signs, $D_{\mathrm{av}}$ is the average GVD\ value, and $\Delta
D\gg D_{\mathrm{av}}$ is the DM amplitude. The size of the DM period is
fixed in Eq. (\ref{Disp_map}) to be $z_{\mathrm{map}}\equiv z_{\mathrm{anom}%
}+z_{\mathrm{norm}}=0.5$ by means of the remaining scaling invariance of Eq.
(\ref{3DGPE_nor}). Below, we focus on the most essential case, with $z_{%
\mathrm{anom}}=z_{\mathrm{norm}}=0.25$.

The multimode character of the system, which resembles GRIN models, may be
demonstrated by expanding the field $u(x,y,\tau ;z)$ into a truncated
superposition of commonly known eigenmodes of the isotropic HO potential in
the $\left( x,y\right) $ plane, while expansion amplitudes are considered as
functions of $\tau $ and $z$, governed by an approximate system of coupled
1D equations. However, we prefer to develop a `` holistic" approach, relying
upon numerical simulations of the fully three-dimensional NLSE (\ref%
{3DGPE_nor}). While the number of actually excited transverse modes is
effectively finite for any 3D solution, the advantage of using the full 3D
equation is that this number is not restricted beforehand. The
applicability of the 3D NLSE for the description of the multimode
propagation in fibers with an internal transverse structure has been
demonstrated, in other contexts, both theopretically and experimentally --
see, e.g., recent work \cite{Wright} and references therein.

As concerns physically relevant scales, the characteristic propagation
length for DM schemes in singe-mode fibers is measured in many kilometers,
while for the multimode waveguides it is limited to a few meters \cite{MM1}-%
\cite{recentMM3}, \cite{sol-MM1}-\cite{we}. This fact makes it possible to
neglect losses in Eq. (\ref{3DGPE}). On the other hand, it may be relevant
to add, to Eq. (\ref{3DGPE_nor}), higher-order terms, representing, in
particular, the third-order GVD and the intra-pulse stimulated Raman
scattering. In this work, we focus on the model based on the basic NLSE (\ref%
{3DGPE}), as it was shown previously that the additional terms, although
affecting the shape of DM solitons, do not lead to dramatic changes in their
dynamics \cite{Frantz,Lakoba,Kaup}.

Basic results for the existence and stability of 3D spatiotemporal solitons
(including ones with intrinsic vorticity), under the action of DM, which are
produced by numerical analysis of the model based on Eqs. (\ref{3DGPE_nor})
and (\ref{Disp_map}), are reported in Section II. Collisions between 3D
solitons in the axial and transverse directions are addressed in Section
III, in the absence and presence of the DM. The paper is concluded by
Section IV.

\section{ Basic numerical results: Stable spatiotemporal solitons under the
action of DM}

\subsection{Families of spatiotemporal solitons in the absence of DM}

First, we produce a family of 3D solitons in the absence of the DM, i.e.,
setting $\Delta D=0$ in Eq. (\ref{Disp_map}). Stationary soliton solutions
to Eq. (\ref{3DGPE_nor}), with real propagation constant $k$, are looked for
as%
\begin{equation}
u(x,y,\tau ;z)=e^{ikz}\phi \left( x,y,\tau \right) ,  \label{k}
\end{equation}%
where the real function $\phi \left( x,y,\tau \right) $ is a localized
solution of equation%
\begin{gather}
k\phi -\frac{1}{2}\left( \frac{\partial ^{2}\phi }{\partial x^{2}}+\frac{%
\partial ^{2}\phi }{\partial y^{2}}\right) -\frac{1}{2}D_{\mathrm{av}}\frac{%
\partial ^{2}\phi }{\partial \tau ^{2}}  \notag \\
+(x^{2}+y^{2})\phi -\phi ^{3}=0.  \label{phi}
\end{gather}%
This equation was solved by means of the Newton conjugate-gradient method
\cite{Yang}. The so obtained soliton solutions are characterized by the
total energy,%
\begin{equation}
E=\int \int \int \left\vert u\left( x,y,\tau \right) \right\vert
^{2}dxdyd\tau ,  \label{E}
\end{equation}%
and the temporal and spatial FWHM widths, $W_{{\tau },\mathrm{FWHM}}$ and $%
W_{{s},\mathrm{FWHM}}$, which are extracted from the numerical data at $%
x=y=0 $ and $\tau =0$, respectively. Then, stability of the solitons was
investigated by means of the linearization of Eq. (\ref{3DGPE_nor}) for
small perturbations, added to the stationary soliton solutions, and
computation of the respective eigenvalues. This was done by means of a
numerical method borrowed from Ref. \cite{Wang}. The so predicted stability
was then verified by direct simulations of Eq. (\ref{3DGPE_nor}).

\begin{figure}[t]
\centerline{\includegraphics[width=3in]{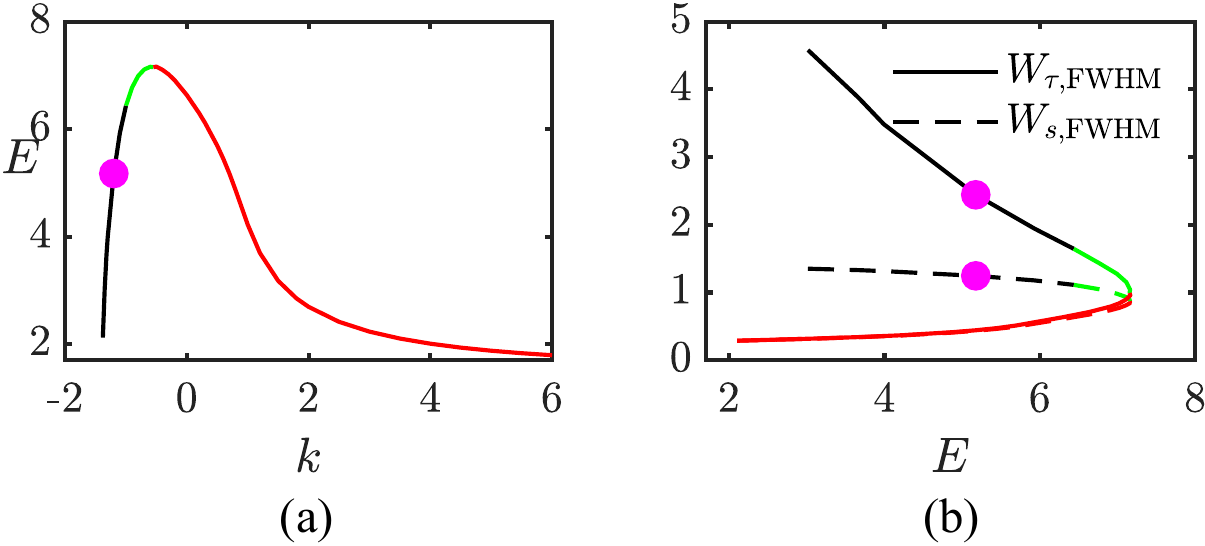}}
\caption{(a) The integral energy (\protect\ref{E}) for the family of
stationary spatiotemporal solitons, obtained in the absence of the DM [$%
\Delta D=0$, $D_{\mathrm{av}}=1$ in Eq. (\protect\ref{Disp_map})], versus
the propagation constant, $k$. The critical point, with $dE/dk=0$, is
located at $k=-0.55$. Solid black, red, and green curves represent,
respectively, stable and unstable stationary solitons, and robust breathers.
(b) The temporal, $W_{{\protect\tau },\mathrm{FWHM}}$, and spatial, $W_{{s},%
\mathrm{FWHM}}$, FWHM widths of the stationary solitons (solid and dashed
lines, respectively) versus $E$. The colors have the same meaning as in (a).}
\label{fig1}
\end{figure}

Note that, in the linear limit and in the absence of GVD, $D_{\mathrm{av}}=0$%
, eigenvalue of $k$ in Eq. (\ref{phi}) corresponds to the commonly known
ground-state energy of the 2D HO potential: $k=-\sqrt{2}$. As seen in Eq. (%
\ref{phi}), the contribution from the anomalous GVD ($D_{\mathrm{av}}>0$) of
temporally localized states shifts $k$ towards more negative values, while
the contribution to $k$ produced by the self-focusing cubic term is
positive, therefore spatiotemporal solitons may exist at both $k<0$ and $k>0$
(as well as at $k=0$), see Fig. \ref{fig1}(a).

The results for the shape and stability of the spatiotemporal solitons are
summarized in Fig. \ref{fig1}, by means of the energy and width curves, $%
E(k) $ and $W_{{\ s,\tau },\text{FWHM}}(E)$, for the soliton family. In
particular, the $E(k)$ dependence in Fig. \ref{fig1}(a) and the stability
boundary demonstrated by it are close to those produced in Ref. \cite{Boris}
for a strongly elongated 3D HO trapping potential, which is close to the 2D
potential in Eq. (\ref{3DGPE_nor}) (in the same work, it was demonstrated
that the $E(k)$ curve can be accurately predicted by means of the
variational approximation). In particular, there are two different solutions
for a given energy, their stability obeying the necessary condition, $%
dE/dk>0 $, i.e., the celebrated Vakhitov-Kolokolov criterion \cite%
{Vakh,Berge}. The stability, as it is shown in Fig. \ref{fig1}, was
identified through the computation of eigenvalues for small perturbations.
In addition to stable and unstable stationary 3D solitons, a narrow interval
of robust breathers, replacing unstable solitons, is found close to the
critical point, $dE/dk=0$ [the short green segment in Fig. \ref{fig1}(a)].
The coexistence of stable and unstable families of 3D solitons, demonstrated
by Fig. \ref{fig1}, is a generic feature of the 3D NLSE with the cubic
self-focusing and HO trapping potential \cite{Boris}.

\begin{figure}[t]
\centerline{\includegraphics[width=3in]{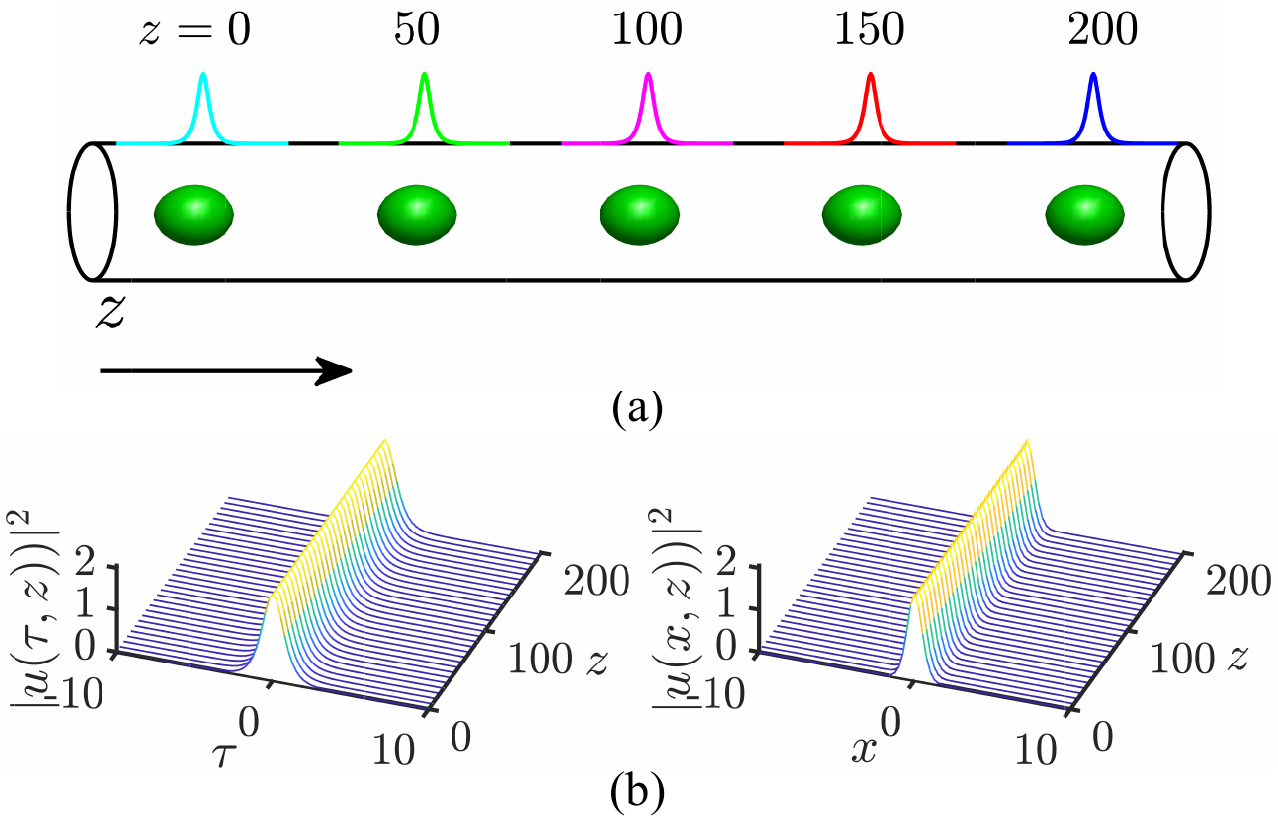}}
\caption{(a) The evolution of the stable spatiotemporal soliton with $k=-1.0$
and energy $E=6.4364$, in the system without DM [$\Delta D=0$, $D_{\mathrm{av%
}}=1$ in Eq. (\protect\ref{Disp_map})], corresponding to the pink dot in
Fig. \protect\ref{fig2}. The evolution is shown by means of the set of
isosurfaces of local intensity, $\left\vert u\left( x,y,\protect\tau \right)
\right\vert ^{2}=0.8$, and its temporal profiles. (b) The evolution of the
temporal (left) and spatial (right) local intensity.}
\label{fig2}
\end{figure}

Results for the stability, displayed in Fig. \ref{fig1}, were verified by
direct simulations of Eq. (\ref{3DGPE_nor}) for the evolution of the
spatiotemporal solitons, performed by means of the split-step
fast-Fourier-transform algorithm. An example, presented in Fig. \ref{fig2}
for the soliton with propagation constant $k=-1$ and energy $E=6.4364$,
corroborates its stability, and displays its 3D shape. The propagation
distance shown in Fig. \ref{fig2} is tantamount to $\sim 200$
dispersion/diffraction lengths of the soliton. In the course of propagation,
the soliton energy is conserved with relative accuracy $\sim 10^{-8}$,
demonstrating that the soliton is an absolutely robust solution of the
underlying DM model. In the absence of DM, the phase chirp of stable
solitons remains equal to zero, in the course of the simulations.

\subsection{Dispersion-managed spatiotemporal solitons}

The main parameter which controls the action of DM in the temporal domain is
the \textit{DM strength} \cite{Nijhof,book,Turitsyn},%
\begin{equation}
S\equiv \left( D_{\mathrm{anom}}z_{\mathrm{anom}}+\left\vert D_{\mathrm{norm}%
}\right\vert z_{\mathrm{norm}}\right) /\left( W_{{\tau },\mathrm{FWHM}%
}^{2}\right) _{\min }~  \label{S}
\end{equation}%
where subscript $\min $ refers to the smallest value of the temporal width
of the periodically oscillating DM soliton. To identify the temporal and
spatial chirps of the soliton oscillating under the action of the DM, $%
C_{\tau }$ and $C_{s}$,\ it is represented in the Madelung's form, as $%
u\left( x,y,\tau ,z\right) \equiv \left\vert u\right\vert \exp \left( i\chi
\right) $. Then, the chirps were computed from the numerically identified
phase, $\chi \left( x,y,\tau ,z\right) $, as%
\begin{gather}
C_{\tau }=\frac{\partial ^{2}}{\partial \tau ^{2}}\chi \left( x=0,y=0,\tau
,z\right) |_{\tau =0},  \notag \\
C_{s}=\frac{\partial ^{2}}{\partial x^{2}}\chi \left( x,y=0,\tau =0,z\right)
|_{x=0},~  \label{C}
\end{gather}

\begin{figure}[t]
\centerline{\includegraphics[width=3in]{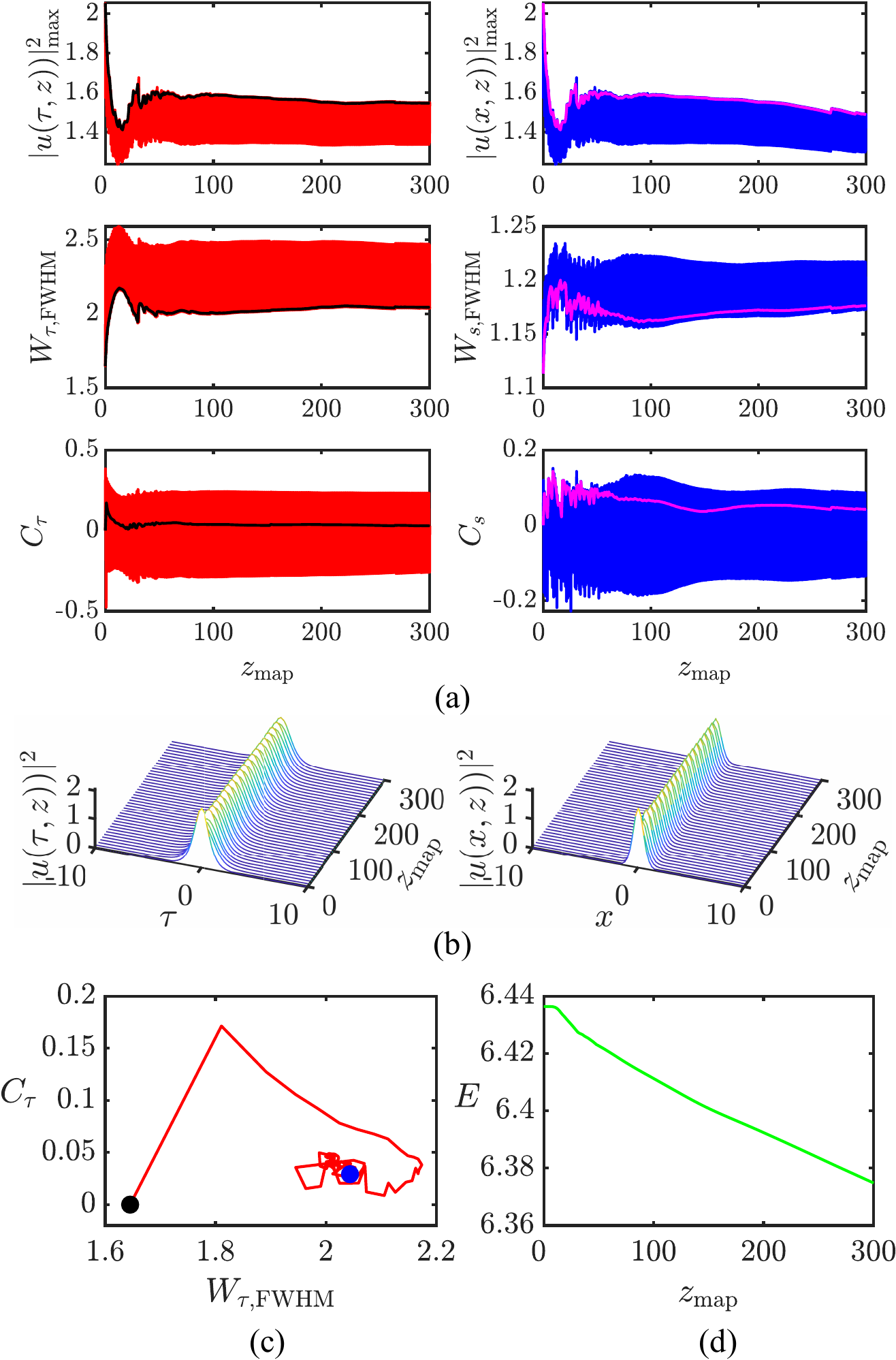}}
\caption{(a) The left column shows the evolution of the peak intensity,
temporal width and chirp of a quasi-stable spatiotemporal soliton under the
action of weak DM with strength $S=0.62$, corresponding to parameters $%
\Delta D=5$ and $D_{\mathrm{av}}=1$ in Eq. (\protect\ref{Disp_map}), over $%
300$ map periods (which corresponds to $z=150$; here and in other figures,
values $z_{\mathrm{map}}$ refer to the propagation distance measured in
units of the DM map). Black lines represent values of the same variables,
taken at the beginning of each DM\ map (\protect\ref{Disp_map}). The input
is the same stationary soliton which is shown in Fig. \protect\ref{fig2} in
the absence of DM, with temporal width $W_{{\protect\tau },\mathrm{FWHM}%
}\approx 1.65$. The right column displays the evolution of the soliton's
spatial width and chirp. Magenta lines represent their values at the
beginning of each DM map. (b) Profiles of the oscillatory solitons in the
temporal (left) and spatial (right) cross sections, plotted at the beginning
of each DM map. (c) The trajectory of the soliton in the plane of the
temporal width ($W_{{\ \protect\tau },\mathrm{FWHM}}$) and chirp ($C_{%
\protect\tau }$). Black and blue markers denote the initial and final points
of the trajectory. (d) The total energy (\protect\ref{E}) of the soliton
versus the propagation distance.}
\label{fig3}
\end{figure}

The simulations of Eq. (\ref{3DGPE_nor}) with the DM map (\ref{Disp_map})
were initiated with an input taken as stable stationary solitons numerically
produced in the system without the DM (see above). It was found that the
outcome of long-distance simulations is adequately determined by the value
of the DM strength (\ref{S}). First, for
\begin{equation}
S<S_{0}\approx 0.93  \label{S0}
\end{equation}%
(relatively weak DM), the simulations produce quasi-stable 3D solitons, as
shown in Fig. \ref{fig3} for $S\approx 0.62$. While the soliton keeps its
overall integrity and does not decay in the course of evolution (Fig. \ref%
{fig3}(d) demonstrates that, having passed $300$ DM\ periods, the soliton
has lost $<1\%$ of the initial energy, through emission of small-amplitude
radiation), it develops quasi-random oscillations of its characteristic
parameters, although with a relatively small amplitude. This is seen in Fig. %
\ref{fig3}(c), that demonstrates a random walk of the soliton's trajectory,
which is trapped in a small domain of the plane of relevant dynamical
parameters, \textit{viz}., the temporal width and chirp.

It is relevant to mention that the spatiotemporal soliton displayed in Fig. %
\ref{fig3}(b) has the dispersion and diffraction lengths $\lesssim 1$. This
is comparable to the underlying DM period, $z_{\mathrm{map}}=0.5$, which
corroborates that the DM is an essential ingredient of the system under
consideration. The same pertains to the completely stable spatiotemporal
soliton displayed below in Fig. \ref{fig4}(b).

The boundary value (\ref{S0}) of the region of quasi-stable DM solitons
corresponds, e.g., to the DM map with $\Delta D\approx 15$ (keeping $D_{%
\mathrm{av}}=1$), and temporal width $\left( W_{{\tau },\mathrm{FWHM}%
}\right) _{\min }\approx 2.84$. At $S>S_{0}$, the propagation of the
spatiotemporal solitons is completely stable under the action of moderate or
strong DM. A typical example is displayed in Fig. \ref{fig4} for $S\approx
1.13$. The most essential manifestation of the full stability is that, in
Fig. \ref{fig4}(d), the input loses $\simeq 7.5\%$ of its total energy at
the initial stage of the evolution ($z<50$ DM periods), adjusting itself to
the propagating state, and then, at $z>50$, the emission of radiation
completely ceases. The fully regular dynamics of the soliton is also
demonstrated by the evolution of its spatial and temporal parameters
displayed in Fig. \ref{fig4}(a), cf. Fig. \ref{fig3}(a) for the quasi-stable
spatiotemporal soliton.

\begin{figure}[t]
\centerline{\includegraphics[width=3in]{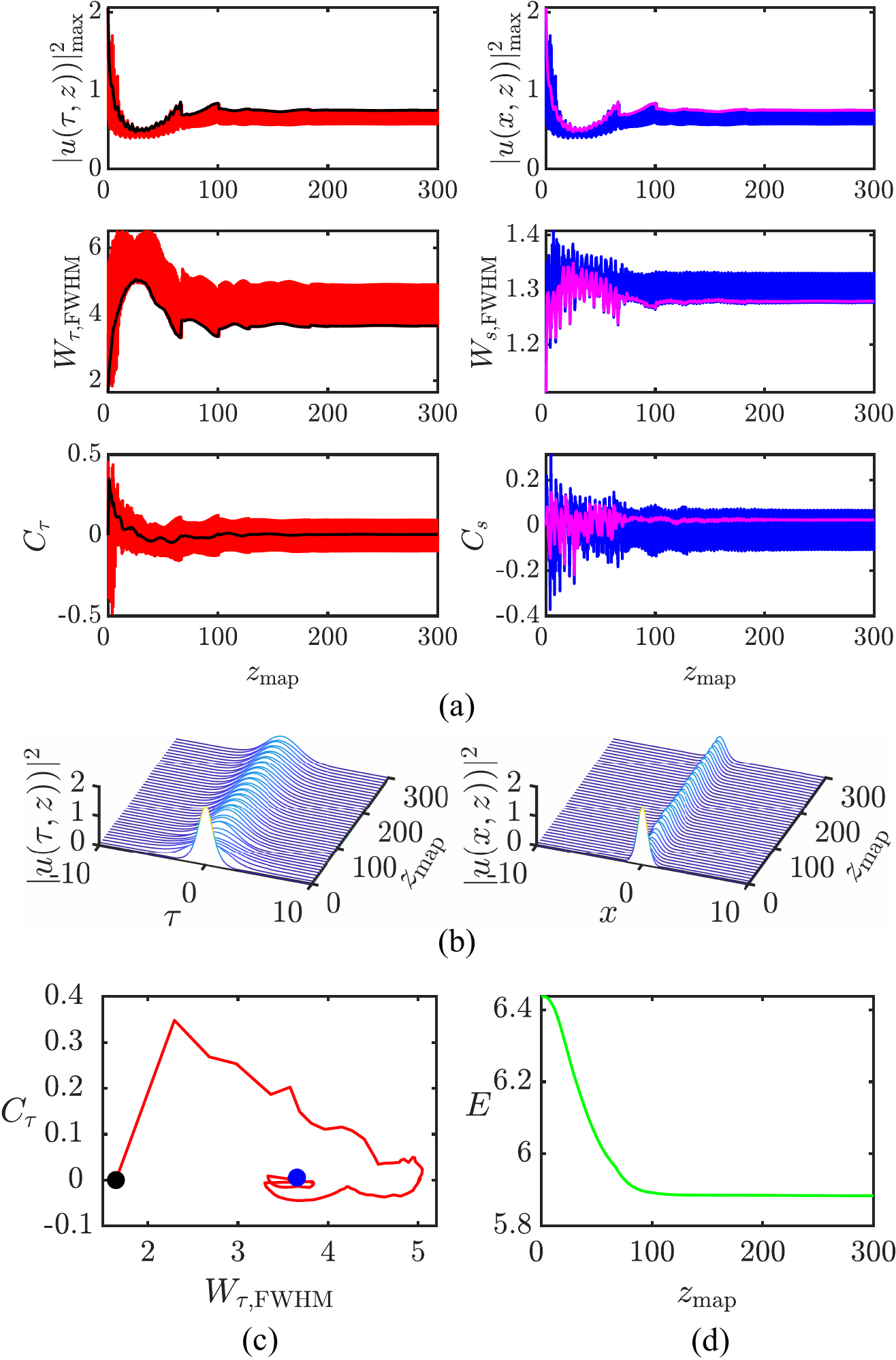}}
\caption{The same as in Fig. \protect\ref{fig3}, and with the same input,
but for the propagation of a stable spatiotemporal soliton under the action
of moderately strong DM, with strength $S\approx 1.13$, corresponding to $%
\Delta D=30$ and $D_{\mathrm{av}}=1$ in Eq. (\protect\ref{Disp_map}). }
\label{fig4}
\end{figure}

\begin{figure}[t]
\centerline{\includegraphics[width=3in]{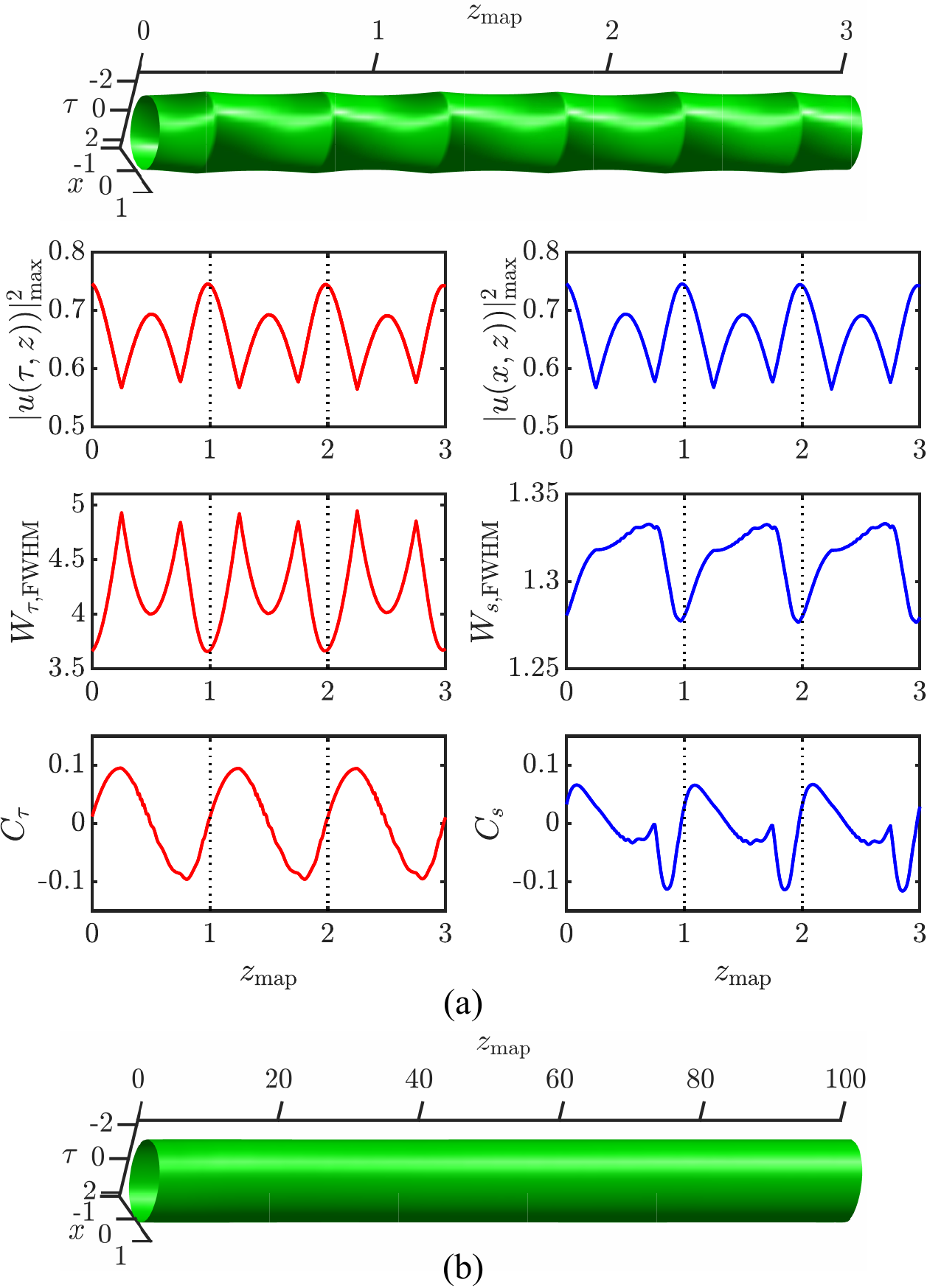}}
\caption{(a) The top panel displays the spatiotemporal dynamics of the
numerically exact DM soliton, with strength $S\approx 1.13$ and minimum
temporal width $\left( W_{{\protect\tau }\mathrm{,FWHM}}\right) _{\min
}\approx 3.65$, to which converges the soliton presented in Fig. \protect\ref%
{fig4}. The evolution of the spatiotemporal DM soliton, produced by means of
the averaging method (see the main text), is displayed by means of an
isosurface of the local intensity, $\left\vert u\left( x,\protect\tau %
,z\right) \right\vert ^{2}=0.25$, in the cross section of $y=0$, as the
soliton passes three DM maps, $0<z<1.5$. The bottom plots show the variation
of the soliton's temporal and spatial characteristics. (b) The isosurface of
$\left\vert u\left( x,0,\protect\tau ,z\right) \right\vert ^{2}=0.25$, shown
at the beginning of each DM map, corroborates the full stability of the
soliton as it passes $100$ DM periods, $0<z<50$.}
\label{fig5}
\end{figure}

Numerically exact DM solitons can be produced by means of the averaging
method, which was previously elaborated for temporal solitons in
dispersion-managed single-mode fibers \cite{Bennion,Nijhof}. The method is
based on collecting a set of shapes of an oscillating soliton, produced by
the straightforward simulations, at points where the shape is narrowest, and
computing an average of the set. The result is the pulse which propagates in
a strictly periodic form. A conclusion of further numerical analysis is that
the direct simulations displayed in Fig. \ref{fig4}, as well as in other
cases of the completely stable propagation, converge precisely to the DM
solitons produced by the averaging method, extended to the present 3D
setting. It is relevant to stress that the averaging procedure needs to be
applied only in the temporal direction, while in the plane of $(x,y)$ the
solution readily converges by itself to the one predicted by the averaging
method. As an example, the DM soliton, to which the evolution displayed in
Fig. \ref{fig4} converges, is shown in Fig. \ref{fig5}. In Fig. \ref{fig5}%
(a), the isosurface plot at the top displays the spatiotemporal evolution of
the DM soliton while it passes three DM maps, $0<z<1.5$, while other panels
display the variation of the soliton's temporal and spatial characteristics.
In Fig. \ref{fig5}(b), the isosurface plot corroborates the full stability
of the soliton over the propagation distance equivalent to $100$ maps.

\subsection{Three-dimensional solitons with embedded vorticity}

The creation of stable spatiotemporal DM solitons with embedded vorticity is
a challenging objective. To the best of our knowledge, 3D solitons of such a
type have not been reported before. To this end, it is necessary, first, to
construct self-trapped vortex states as solutions to Eq. (\ref{3DGPE_nor})
in the absence of DM ($\Delta D=0$). In terms of the polar coordinates, $%
\left( r,\theta \right) $ in the plane of $\left( x,y\right) $, stationary
vortex solitons are sought for as%
\begin{equation}
u=\exp \left( im\theta \right) \phi (r,\tau ),  \label{psi}
\end{equation}%
with integer vorticity $m\geq 1$ and a localized real function $\phi $
obeying equation%
\begin{gather}
k\phi -\frac{1}{2}\left( \frac{\partial ^{2}}{\partial r^{2}}+\frac{1}{r}%
\frac{\partial }{\partial r}-\frac{m^{2}}{r^{2}}-\frac{1}{2}D_{\mathrm{av}}%
\frac{\partial ^{2}}{\partial \tau ^{2}}\right) \phi  \notag \\
+(x^{2}+y^{2})\phi -\phi ^{3}=0,  \label{psipsi}
\end{gather}%
with boundary condition $\phi \sim r^{m}$ at $r\rightarrow 0$. In the linear
limit and for $D_{\mathrm{av}}=0$, $k$ is determined by energy eigenvalues
of the 2D HO potential, i.e., $k_{\mathrm{linear}}=-\left( 1+m\right) \sqrt{2%
}$.

\begin{figure}[t]
\centerline{\includegraphics[width=3in]{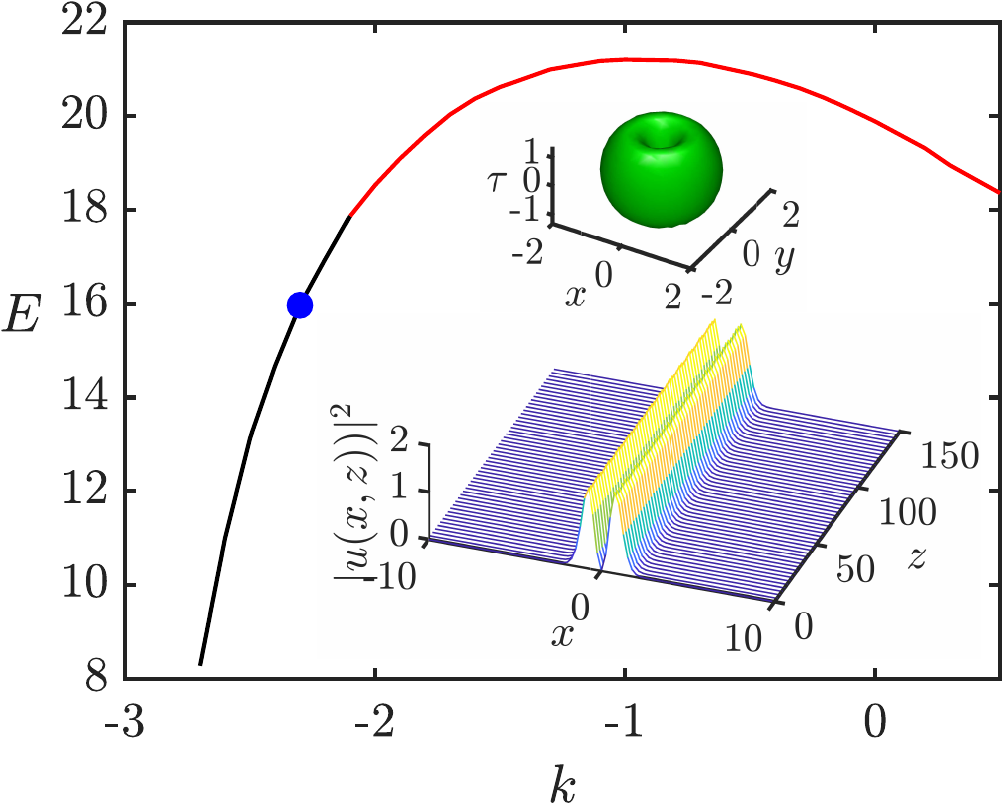}}
\caption{The energy of numerically generated vortex-soliton solutions of Eq.
(\protect\ref{3DGPE_nor}) with $D_{\mathrm{av}}=1$ and $m=1$ [see Eq. (%
\protect\ref{psi})], in the absence of DM ($\Delta D=0$), vs. the
propagation constant. Black and red segments denote stable and unstable
subfamilies, as identified through the computation of eigenvalues for small
perturbations. The top inset shows the isosurface plot, $|u(x,y,\protect\tau %
)|^{2}=0.35$, of a typical stable vortex soliton corresponding to the dot on
the $E(k)$ curve, with $E=15.97$ and $k=-2.3$. The bottom inset: the
stability of this soliton is illustrated by simulations of its perturbed
evolution, displayed on the line of $y=\protect\tau =0$. An example of
unstable evolution of the vortex soliton, corresponding to $k=0$, is
presented in Fig. \protect\ref{fig7}.}
\label{fig6}
\end{figure}

\begin{figure}[t]
\centerline{\includegraphics[width=0.9\columnwidth]{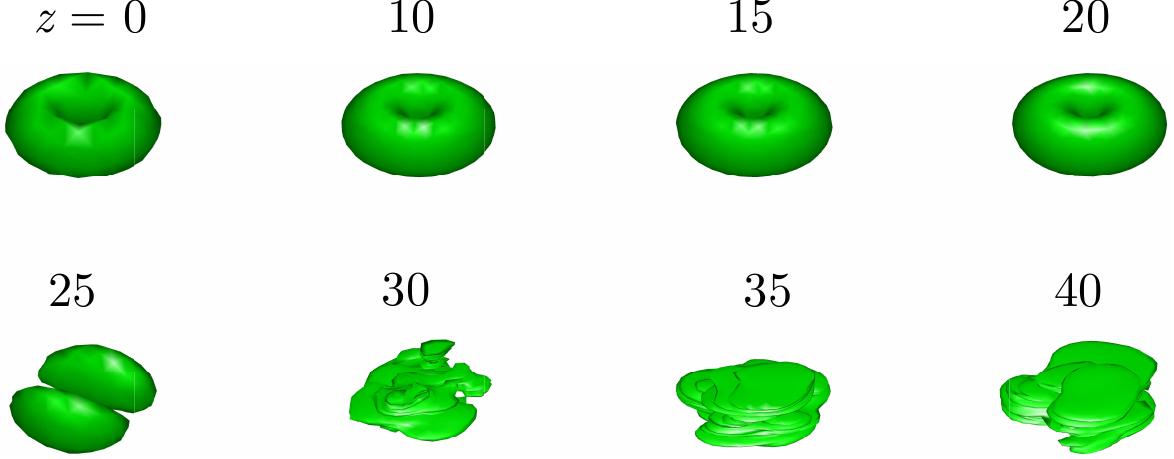}}
\caption{The isosurface evolution of an unstable vortex soliton with $m=1$,
simulated in the framework of Eq. (\protect\ref{3DGPE_nor}) with $\Delta D=0$
and $D_{\mathrm{av}}=1$. The input is a stationary solution of Eq. (\protect
\ref{psipsi}) for $k=0$, which is definitely unstable, as seen in Fig.
\protect\ref{fig6}.}
\label{fig7}
\end{figure}

A family of vortex-soliton solutions to Eq. (\ref{psipsi}) was constructed
by means of the same Newton conjugate-gradient method which was used above
for producing fundamental solitons, and the stability was explored through
the computation of eigenvalues for modes of small perturbations. The result
for $m=1$, similar to that reported in Ref. \cite{Boris} for the NLSE with a
strongly anisotropic HO trapping potential (i.e., the Gross-Pitaevskii
equation), is displayed in Fig. \ref{fig6}. As seen in the figure, the
Vakhitov-Kolokolov criterion, $dE/dk>0$, is necessary but not sufficient for
the stability of the vortex solitons. An example of the stable vortex, shown
by means of the isosurface of $\left\vert u\left( x,y,\tau \right)
\right\vert ^{2}$, clearly shows the inner hole, maintained in the soliton
by the embedded vorticity. Unstable vortices follow the usual scenario,
spontaneously splitting in a pair of fragments \cite{PhysD-vortices}, as
shown in Fig. \ref{fig7}. Eventually, the fragments merge into a
quasi-turbulent state.

\begin{figure}[t]
\centerline{\includegraphics[width=3in]{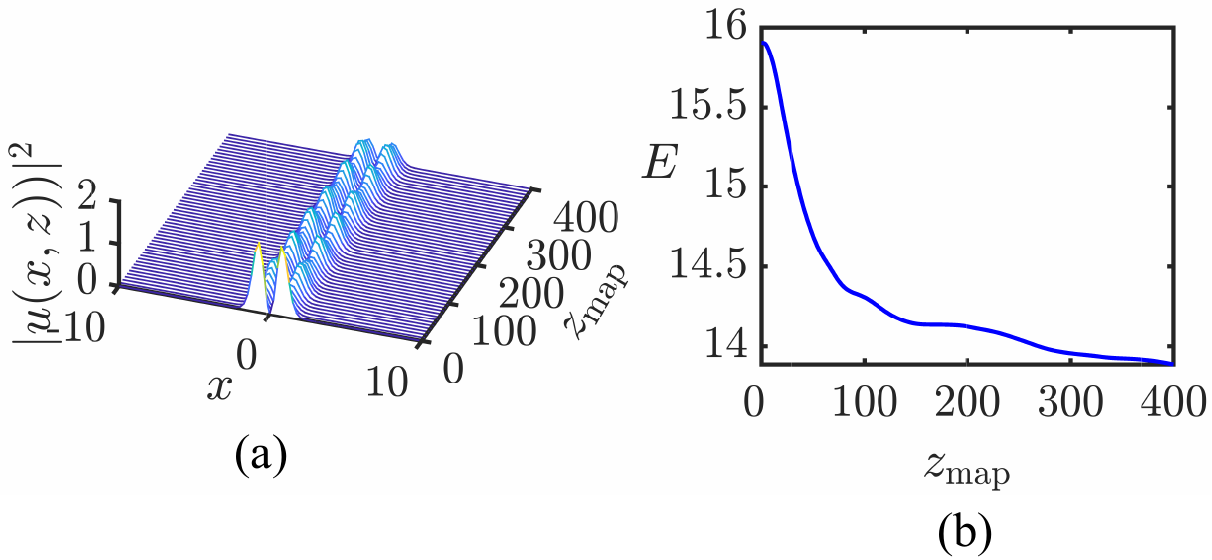}}
\caption{(a) Quasi-stable evolution of a vortex soliton with $m=1$,
simulated in the framework of Eq. (\protect\ref{3DGPE_nor}) with relatively
strong DM, \textit{viz}., $\Delta D=30$ and $D_{\mathrm{av}}=1$. The input
is the stationary vortex soliton with $\Delta D=0$ and $D_{\mathrm{av}}=1$
shown in Fig. \protect\ref{fig6}. (b) The total energy of the vortex soliton
versus the propagation distance.}
\label{fig8}
\end{figure}

In the presence of the DM, quasi-stable vortex solitons were found by direct
simulations. A typical example is shown in Fig. \ref{fig8}: using a stable
vortex from Fig. \ref{fig6}, which was found for $\Delta D=0$, as the input,
direct simulations demonstrate the formation of a robust soliton which keeps
the intrinsic vorticity and features slow shape oscillations with a large
period, $\Delta z\simeq 80$ DM periods, under the action of the relatively
strong DM, with $\Delta D=30$. The shape oscillations may be removed by
accurately tuning the input, and a stability area for vortex solitons may be
thus identified in the parameter space. Here, we do not aim to report such
results in a comprehensive form, as it is extremely time-consuming to
accumulate the necessary amount of numerical data. We did not consider
vortex solitons with $m>1$ either, as it is known that, under the action of
the 3D HO trapping potential, 3D solitons with multiple embedded vorticity
are completely unstable, even in the absence of DM \cite%
{Boris,PhysD-vortices}.

\section{Collisions between three-dimensional solitons}

\subsection{Collisions in the longitudinal direction}

Once stable DM\ solitons have been found, it is relevant to consider
collisions between them. The DM soliton can be set in longitudinal motion by
the application of a kick (i.e., longitudinal boost) to it, with arbitrary
frequency shift $\Omega $. Indeed, Eq. (\ref{3DGPE_nor}) is invariant with
respect to the Galilean transformation, which generates a new solution from
a given one, $u_{0}\left( x,y,\tau ;z\right) $:%
\begin{gather}
u\left( x,y,\tau ;z\right) =\exp \left( -i\Omega \tau -\frac{i}{2}\Omega
^{2}\int D(z)dz\right)  \notag \\
\times u_{0}\left( x,y,\tau -\tau _{0}+\Omega \int D(z)dz;z\right) ,
\label{u}
\end{gather}%
where $\tau _{0}$ is an arbitrary constant shift of the solution as a whole.
Note that, in addition to the progressive motion with average speed
\begin{equation}
\mathrm{speed}=-D_{\mathrm{av}}\Omega ,  \label{sp}
\end{equation}%
the substitution of DM map (\ref{Disp_map}) in Eq. (\ref{u}) gives rise to
oscillatory motion with spatial period $z_{\mathrm{map}}=0.5$ and temporal
amplitude
\begin{equation}
\Delta \tau =(1/2)\Delta D|\Omega |z_{\mathrm{map}}~.  \label{Delta}
\end{equation}

Thus, it is possible to create initial conditions for simulating collisions
between spatiotemporal DM solitons moving in opposite temporal directions.
In studies of temporal DM solitons in single-mode optical fibers, collisions
were studied in detail for solitons carried by different wavelengths in the
wavelength-division-multiplexed (WDM) system, which is an important
practical problem, as the collision-induced jitter is a source of errors in
data-transmission schemes \cite{coll1,coll2}.

\begin{figure}[t]
\centerline{\includegraphics[width=3in]{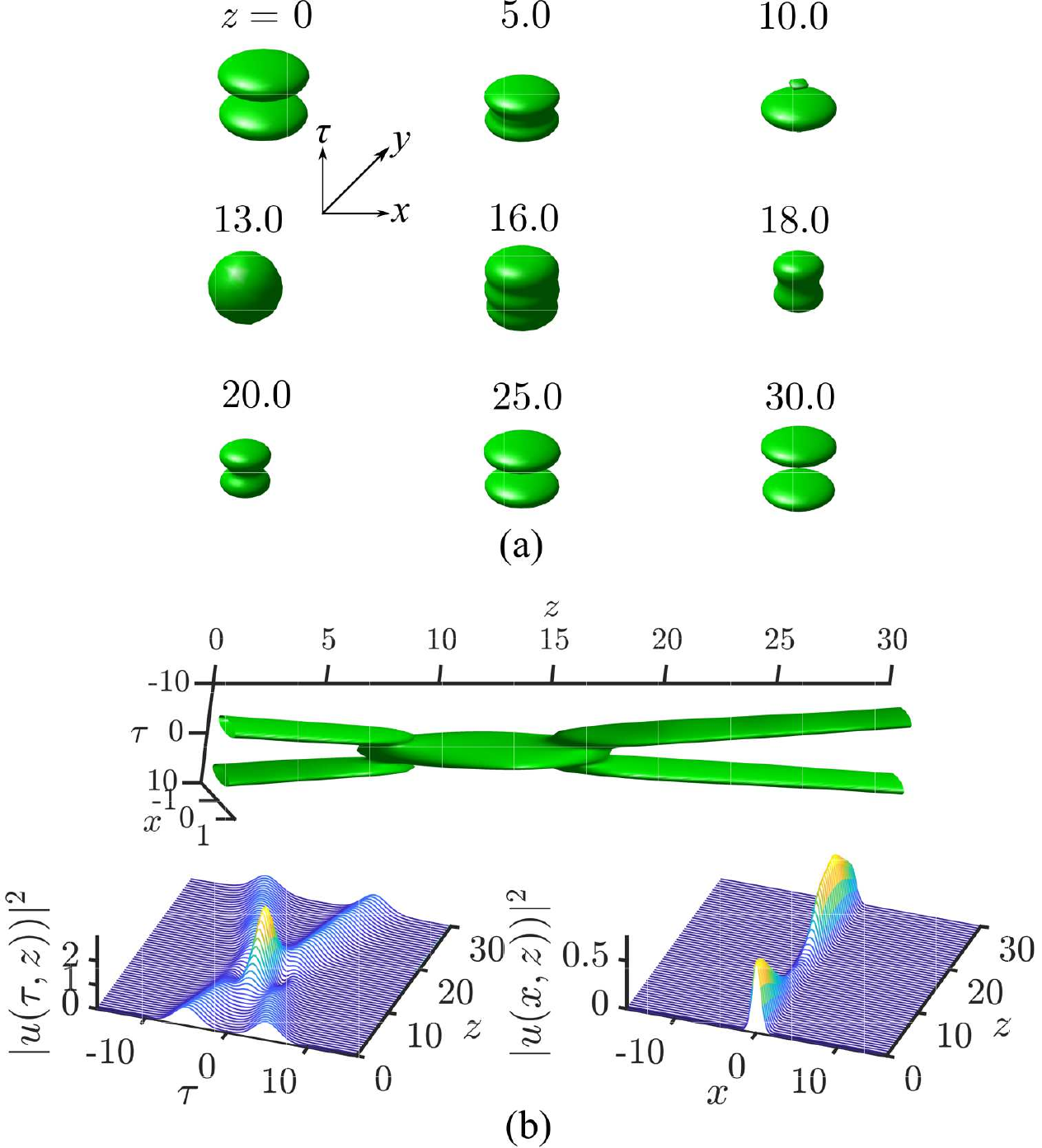}}
\caption{Simulations of slow collision, in the axial (temporal) direction,
between two DM solitons, constructed as per Eq. (\protect\ref{u}) from the
stable DM mode produced in Fig. \protect\ref{fig4} [with $\Delta D=30$, $D_{%
\mathrm{av}}=1$, $\left( W_{{\protect\tau },\mathrm{FWHM}}\right) _{\min
}\approx $ $3.65$]. The solitons, initially set at positions $\protect\tau %
_{0}=\pm 5$, are boosted by frequency shifts $\Omega =\pm 0.3$. (a) A set of
3D snapshots, taken in the course of the collision, are depicted by means of
isosurfaces of the local intensity, $\left\vert u\left( x,y,\protect\tau %
\right) \right\vert ^{2}$. (b) The collision shown by means of the
isosurface in the cross section of $y=0$ (the top plot), and by the
evolution of the intensity on the lines of $(x=y=0)$ and $(y=\protect\tau =0)
$ (the left and right bottom plots, respectively).}
\label{fig9}
\end{figure}

Numerical simulations demonstrate elastic collisions between boosted
solitons, created as per Eq. (\ref{u}), with frequency shifts $\pm |\Omega |$
and centers initially set at points $\tau =\pm \tau _{0}$, see an example in
Fig. \ref{fig9} for $\Omega =\pm 0.3$. This `` slow" collision is strongly
affected by the DM, because, with the smallest value of the solitons' width $%
\left( W_{{\ \tau },\mathrm{FWHM}}\right) _{\min }\approx $ $3.65$ and
relative speed $2D_{\mathrm{av}}|\Omega |=0.6$ [see Eq. (\ref{sp})], the
completion of the collision requires passing the propagation distance which
is tantamount to more than $10$ DM periods: $Z_{\mathrm{coll}}\approx
2\left( W_{{\tau },\mathrm{FWHM}}\right) _{\min }/\left( 2|\Omega |\right)
\simeq 12z_{\text{\textrm{map}}}$. This estimate implies that multiple
collisions take place between the two solitons, while they stay strongly
overlapping. Indeed. Eq. (\ref{Delta}) yields an estimate for the
characteristic overlapping degree: $\left( W_{{\tau },\mathrm{FWHM}}\right)
_{\min }/\Delta \tau \approx 1.62$ for the present values of parameters.
This circumstance explains a complex intermediate structure observed at $z=16
$ in Fig. \ref{fig9}(a), and in the left bottom plot of Fig. \ref{fig9}(b).

In addition to the initial temporal separation $2\tau _{0}$, a pair of
colliding solitons is characterized by a phase shift between them, $\Delta
\chi $. However, additional numerical results clearly demonstrate that,
similar to what is well known in many other systems, results of the
collisions do not depend on $\Delta \chi $. Actually, the collisions are
driven by $\Delta \chi $ in the case when the initial pair is taken with
zero relative velocity and relatively small separation \cite{Hulet}, which
is not the case here. Nevertheless, collisions displayed below in Fig. \ref%
{fig11} are sensitive to the fact they start with $\Delta \chi =0$.

It is worthy to note that the collision produces, in a short interval of the
propagation distance, a peak with a large amplitude, as seen in the left
bottom panel of Fig. \ref{fig9}(b). This picture somewhat resembles the
formation of optical rogue waves, see, e.g., recent work \cite{rogue} and
references therein. However, the analogy is rather superficial, as, unlike
traditional rogue waves, here the peak appears not spontaneously, but as a
result of the collision, it is not fed by a continuous-wave background, and
the configuration is not intrinsically unstable (as it does not include a
modulationally unstable background).

\subsection{Collisions in the transverse direction}

In the present context, it is relevant to consider, as a new option,
collisions between 3D solitons moving in the transverse direction in the 3D
setting. For this purpose, solitons may be set in motion by initially
placing them at off-axis positions, with nonzero initial coordinates of the
soliton's centers, $\left( x_{0},y_{0}\right) $, and letting them roll down
towards $x=y=0$ in the HO potential (a similar setup was employed to
initiate collisions between matter-wave solitons trapped in the 3D isotropic
HO potential \cite{Hulet}).

\begin{figure}[t]
\centerline{\includegraphics[width=3in]{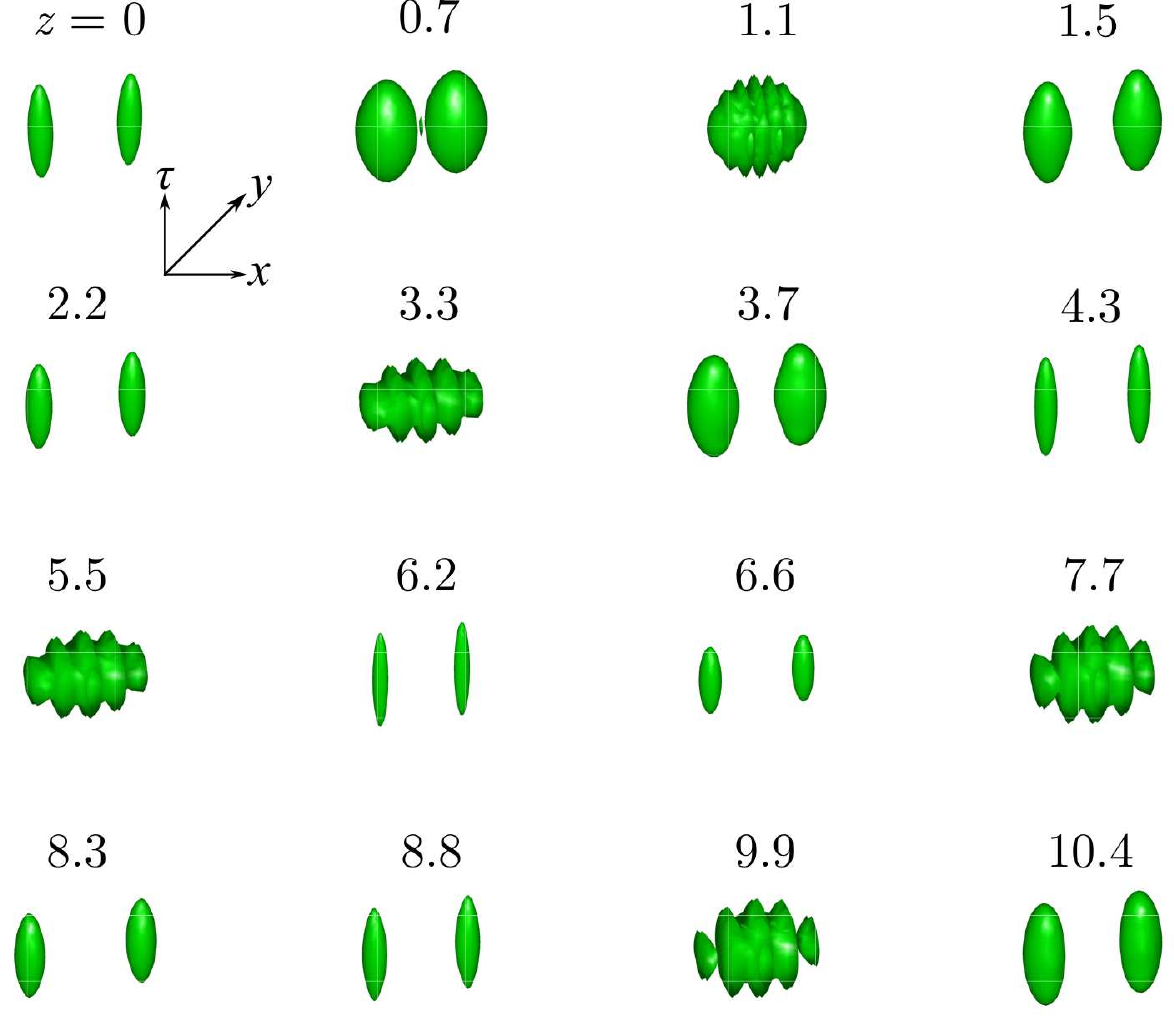}}
\caption{Collisions, in the transverse plane, between two 3D solitons in the
absence of DM, i.e., with $\Delta =0$ and $D_{\mathrm{av}}=1$ in Eqs. (%
\protect\ref{psipsi}) and (\protect\ref{Disp_map}). At $z=0$, the solitons,
identical to the one presented in Fig. \protect\ref{fig2}, are created at
off-axis positions $x_{0}=y_{0}=\pm 3$. The dynamics of the collision is
shown by means of the set of isosurfaces of the local intensity, $\left\vert
u\left( x,y,\protect\tau \right) \right\vert ^{2}$. The quasi-elastic
collisions between the solitons repeat periodically in the course of
indefinitely long simulations.}
\label{fig10}
\end{figure}

Thus, it is possible to consider the collision between two identical
solitons initially shifted to positions $\pm \left( x_{0},y_{0}\right) $. It
makes sense to address this possibility, first, in the absence of DM ($%
\Delta D=0$), as it was not addressed in previous works. As shown by a
typical simulation displayed in Fig. \ref{fig10}, the colliding solitons,
originally created (at $z=0$) at positions $x_{0}=y_{0}=\pm 3$, pass through
each other quasi-elastically (QE) at $z\approx 1.1$, separate, then return
under the action of the transverse HO potential, and again feature the QE
collision at $z\approx 3.3$. The extension of the simulation demonstrates
that the chain of QE collisions between the solitons, which perform the
shuttle motion in the HO potential, continues indefinitely long, with
intervals
\begin{equation}
\Delta z_{\mathrm{coll}}\approx 2.2  \label{2.2}
\end{equation}%
between the collisions. Note that, with the frequency of the trapping HO\
potential $\omega _{\perp }=\sqrt{2}$ in Eq. (\ref{psipsi}), the half-period
of the shuttle motion in the HO potential is $\Delta z_{\mathrm{oscill}}=\pi
/\omega _{\perp }\approx \allowbreak 2.22$, which readily explains the size
of the interval between the collisions, see Eq. (\ref{2.2}). Similar results
were produced by simulations of head-on collisions between a soliton
released from the initial position with coordinates $x_{0}=y_{0}>0$ and its
quiescent counterpart placed on-axis (at $x_{0}=y_{0}=0$). Similar to what
is mentioned above for collisions in the longitudinal direction, head-on
collisions in the transverse plane are not sensitive to $\Delta \chi $
either.

\begin{figure}[t]
\centerline{\includegraphics[width=3in]{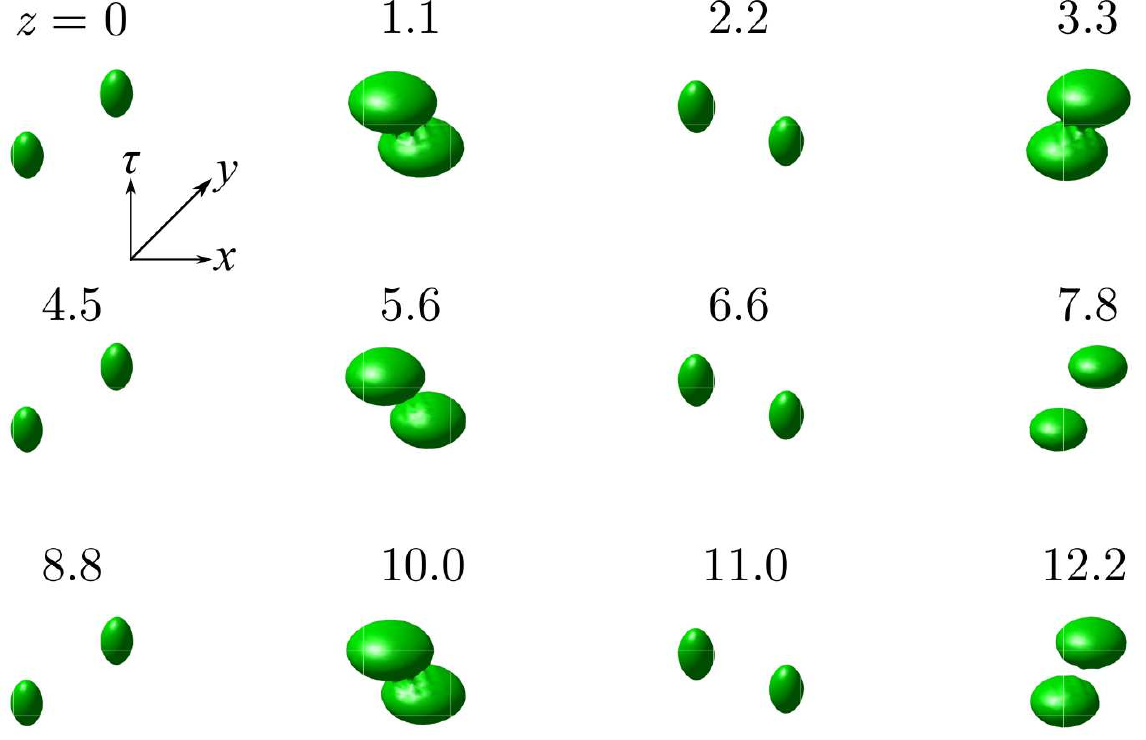}}
\caption{The same as in Fig. \protect\ref{fig10}, but in the case when
centers of the two in-phase solitons are initially separated in the axial
direction by shifts $\protect\tau _{0}=\pm 3$.}
\label{fig11}
\end{figure}

It is also relevant to consider interactions between two solitons initially
created with centers placed at points with coordinates $\left( x,y\right)
=\pm \left( x_{0},y_{0}\right) $, additionally separated in the axial
direction, i.e., with temporal coordinates $\tau =\pm \tau _{0}$. In this
case, the solitons do not collide head-on; nevertheless, the simulation
displayed in Fig. \ref{fig11} demonstrates that the attractive interaction
between the in-phase 3D solitons, mediated by their tails \cite{attraction},
leads to the collision between them. As mentioned above, these collisions,
unlike those displayed in Figs. \ref{fig9} and \ref{fig10}, are sensitive to
the fact that the solitons were created with zero initial phase difference, $%
\Delta \chi =0$; in the case of $\Delta \chi =\pi $, the collision does not
take place, as the solitons interact repulsively (not shown here in detail).
In this case, simulations also demonstrate a periodic chain of QE collisions
with the same interval as predicted by Eq. (\ref{2.2}). However, the
corresponding full period of the collisional dynamics is double (including
two collisions), $\Delta z_{\mathrm{coll}}^{\mathrm{(full)}}\approx 4.4$,
because each collision leads to rotation of the line connecting centers of
the solitons, as seen in Fig. \ref{fig11}.

\begin{figure}[t]
\centerline{\includegraphics[width=3in]{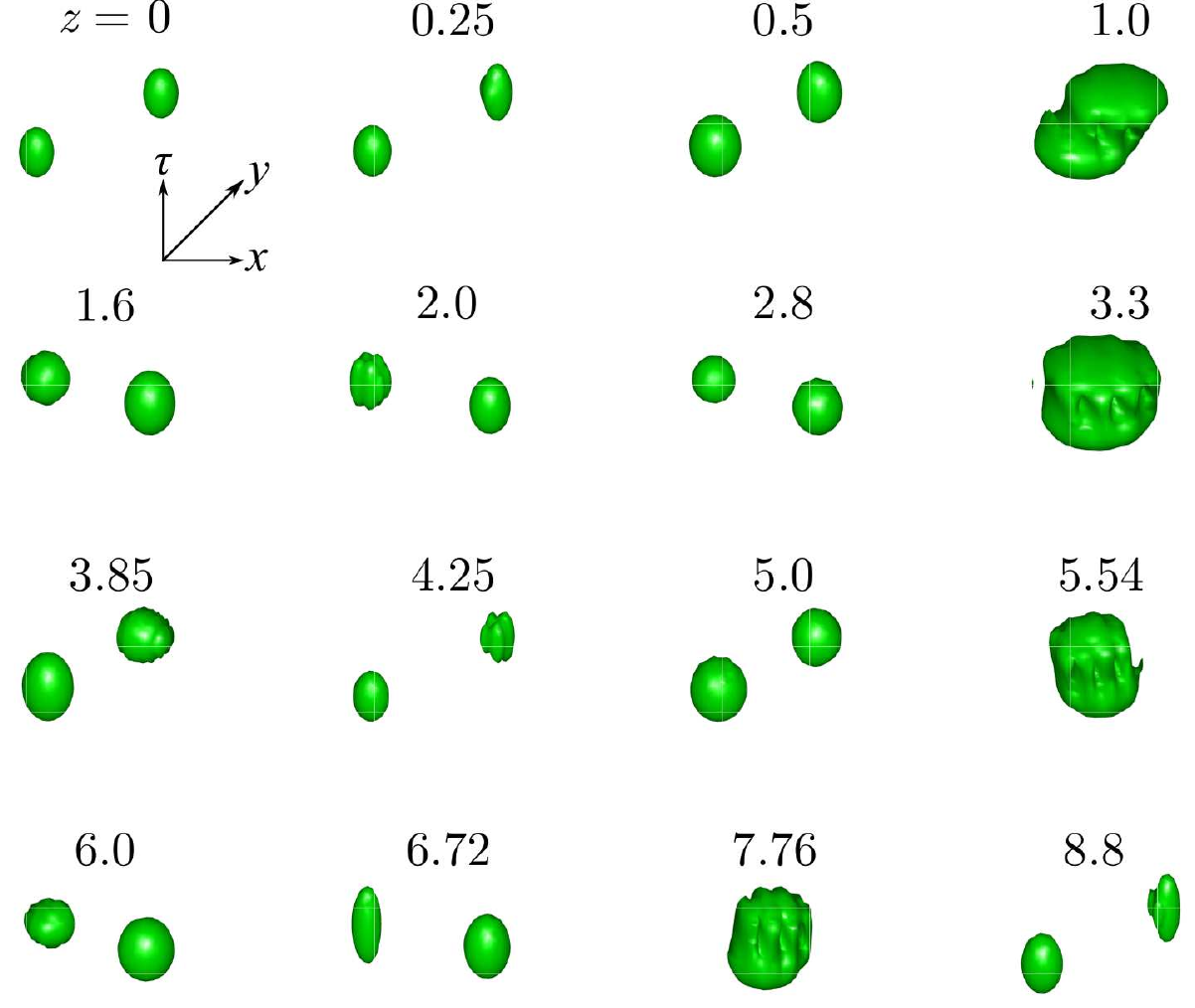}}
\caption{Collisions, in the transverse plane, between two 3D solitons in the
presence of DM, with $\Delta D=30$ and $D_{\mathrm{av}}=1$ in Eqs. (\protect
\ref{psipsi}) and (\protect\ref{Disp_map}). At $z=0$, a DM soliton, the same
as the one presented in Figs. \protect\ref{fig4} and \protect\ref{fig5}, is
placed at the position with $x_{0}=y_{0}=5$, and an identical soliton is
placed at $x_{0}=y_{0}=0$. The dynamics of the collision is shown by means
of the set of isosurfaces of the local intensity, $\left\vert u\left( x,y,%
\protect\tau \right) \right\vert ^{2}$. Quasi-elastic collisions between the
solitons recur periodically in the course of the evolution.}
\label{fig12}
\end{figure}

In the presence of the DM, collisions in the transverse direction are
similar to those shown in Figs. \ref{fig10} and \ref{fig11}. In particular,
Fig. \ref{fig12} demonstrates simulations of the head-on collision between
the DM soliton, released from the position with $x_{0}=y_{0}=3$, and a
quiescent one, placed at $x_{0}=y_{0}=0$. Note that, at the respective
values of the parameters, the transverse speed of the soliton rolling down
under the action of the HO potential at the collision moment is $\left(
\mathrm{speed}\right) _{\perp }=-\omega _{\perp }\sqrt{x_{0}^{2}+y_{0}^{2}}%
=-5\sqrt{2}$. Then, the propagation distance necessary for the completion of
the collision in the transverse direction can be estimated as%
\begin{equation}
\left( \Delta z_{\mathrm{coll}}\right) _{\perp }\simeq 2W_{{\ s},\mathrm{FWHM%
}}/\left\vert \left( \mathrm{sp}\right) _{\perp }\right\vert \simeq 0.4,
\label{Delta2}
\end{equation}%
where estimate $W_{{s},\mathrm{FWHM}}\simeq 1.3$ is taken from Fig. \ref%
{fig4}(b). The comparison of the value given by Eq. (\ref{Delta2}) with $z_{%
\mathrm{map}}=0.5$ suggests that the DM produces a moderate effect on the
collisions, as corroborated by the comparison of Figs. \ref{fig10} and \ref%
{fig12}. In any case, the collisions keep their QE character, recurring
periodically with the interval correctly predicted by Eq. (\ref{2.2}).

\begin{figure}[t]
\centerline{\includegraphics[width=3in]{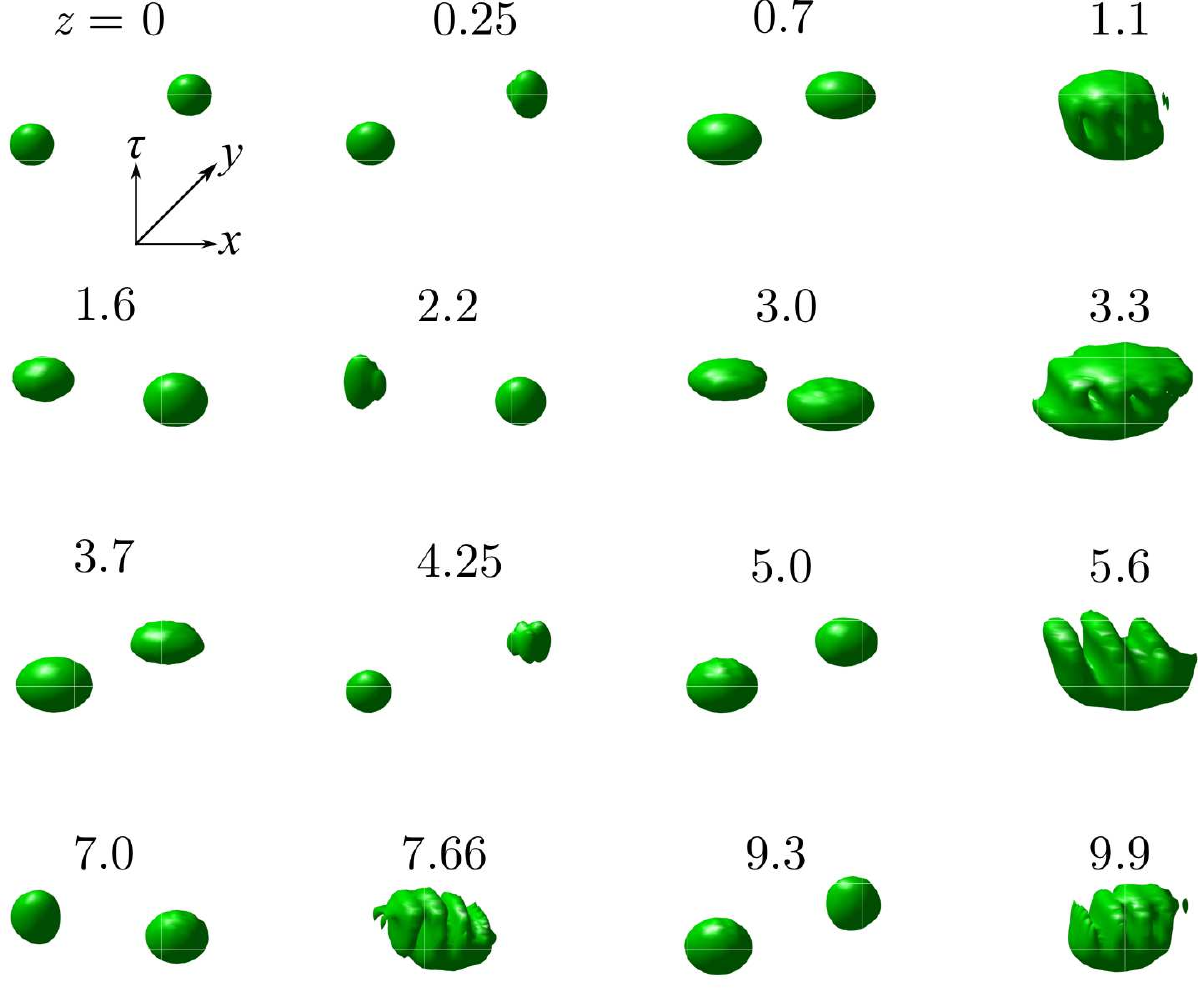}}
\caption{Recurring collisions between the same DM\ solitons as in Fig.
\protect\ref{fig12}, but initially placed at positions with coordinates $%
x_{0}=y_{0}=\protect\tau _{0}=5$ and $x_{0}=y_{0}=\protect\tau _{0}=0$,
i.e., with the additional initial separation $\Delta \protect\tau =5$ in the
axial (temporal) direction.}
\label{fig13}
\end{figure}

As shown in Fig. \ref{fig13}, two DM solitons, initially taken with an
additional temporal (axial) separation between them (therefore they cannot
collide head-on), collide too, due to the mutual attraction, similar to what
is shown above in Fig. \ref{fig11} for the 3D solitons in the absence of DM.
The interval between consecutive collisions is accurately predicted, as
above, by Eq. (\ref{2.2}), while, similar to what is observed in Fig. \ref%
{fig11}, the full period of the collisional dynamics is double, $\Delta z_{%
\mathrm{coll}}^{\mathrm{(full)}}\approx 4.4$, including two collisions, as
the line connecting centers of the interacting solitons rotates, as a result
of each collision. The comparison of Figs. \ref{fig11} and (\ref{fig13})
suggests that, as well as in the case of the head-on collision, DM produces
a moderate effect on the interaction of the spatiotemporal solitons.

\section{Conclusion}

The objective of this work is to extend the concept of DM (dispersion
management) for 3D spatiotemporal solitons in multimode nonlinear optical
fibers. In previous works, the DM concept was elaborated in detail,
theoretically and experimentally, for 1D temporal solitons in single-mode
fibers, as well as, in a theoretical form, for 2D spatiotemporal `` light
bullets" in planar waveguides. We have here produced a family of 3D solitons
in the model combining the GRIN structure of the refractive index in the
transverse plane, approximated by the HO (harmonic-oscillator, i.e.,
quadratic) profile, Kerr self-focusing nonlinearity, and the usual DM map,
based on periodic alternation of anomalous- and normal-GVD segments. It is
found that the stability of the spatiotemporal DM solitons is determined by
the DM\ strength parameter, $S$: periodically oscillating DM\ solitons
self-trap from localized inputs, as fully stable modes, at $S>S_{0}\simeq
0.93$. After a transient rearrangement of the input, stable DM solitons keep
constant energy. At $S<S_{0}$, the simulations demonstrate quasi-stability:
the self-trapping gives rise to spatiotemporal solitons with persistent
small-amplitude random intrinsic vibrations, which very slowly lose their
energy, remaining robust over propagation distances corresponding to
hundreds of DM periods. Stable three-dimensional DM solitons with embedded
vorticity were constructed too. Collisions between DM solitons were
considered by boosting them in opposite axial directions, with a conclusion
that the collisions are quasi-elastic. Collisions in the transverse plane
were also addressed, by initially placing one or both solitons at off-axis
positions, and letting them roll down under the action of the HO potential.
In this case, the pair of solitons, which perform the shuttle motion in the
confining potential, demonstrate a periodic sequence of quasi-elastic
collisions, in the absence or presence of DM.

The present work can be extended by considering the system including DM
combined with the nonlinearity management. This extension may be relevant
because different segments of multimode fibers in the composite waveguide
may have different values of the nonlinearity parameter. As mentioned above,
the realistic DM model may also include the intra-pulse
stimulated-Raman-scattering effect and third-order dispersion, which
deserves the consideration. Another direction for the development of the
analysis may be the systems of the WDM type, with two or several distinct
carrier wavelengths, modeled by a system of nonlinearly coupled NLSEs.

\section{Acknowledgments}

We appreciate valuable discussions with T. Birks and L. G. Wright. This work
was supported, in part, by the Thailand Research Fund (grant No. BRG
6080017), Israel Science Foundation (grant No. 1286/17), and Russian
Foundation for Basic Research (grant No. 17-02-00081). TM  acknowledges the
support from Faculty of Engineering, Naresuan University, Thailand.

\end{document}